\documentclass{NIME-alternate}
\usepackage{url}
\begin{document}
\CopyrightYear{2010, }
\crdata{Copyright remains with the author(s).}
\conferenceinfo{NIME2010, June 15-18, 2010,}{Sydney, Australia}
\title{Cross-artform performance using networked interfaces: \ttlit{Last Man to Die's Vital LMTD}}
\numberofauthors{3}
\author{
\alignauthor Charles Martin\\
     \affaddr{Last Man to Die}\\
       \affaddr{Canberra, Australia}\\
       \email{cpm@charlesmartin.com.au}
\alignauthor Benjamin Forster\\
     \affaddr{Last Man to Die}\\
       \affaddr{Canberra, Australia}\\
       \email{benjamin@\\emptybook.net}
\alignauthor Hanna Cormick\\
       \affaddr{Last Man to Die}\\
       \affaddr{Canberra, Australia}\\
       \email{hcormick@gmail.com}
}

\date{19 April 2010}
\maketitle

\begin{abstract}
In 2009 the cross artform group, \textit{Last Man to Die}, presented a series of performances using new interfaces and networked performance to integrate the three artforms of its members (actor, Hanna Cormick, visual artist, Benjamin Forster and percussionist, Charles Martin). This paper explains our artistic motivations and design for a computer vision surface and networked heartbeat sensor as well as the experience of mounting our first major work, \textit{Vital LMTD}.
\end{abstract}


\section{Introduction}

\noindent The cross artform group, \textit{Last Man to Die}~\cite{LastManToDie-website} was founded in 2008 by Charles Martin (percussionist), Benjamin Forster (visual artist) and Hanna Cormick (actor). Our goal is to create works where our three disciplines influence each other implicitly through novel interfaces. 

For Martin and Forster, computer interfaces are a vital part of their artistic practice. Martin's work is focussed on augmenting percussion instruments as well as using novel interfaces with percussive techniques. Forster had worked with generative art using a physical plotter to draw as well as projection art. Cormick (actor) is a specialist in impulse movement and mask.

Our first major work \textit{Vital LMTD}~\cite{LastManToDie-VitalLMTD} was presented at \textit{The Street Theatre}, Canberra in October 2009. This work was a series of set-pieces exploring different interactions between our artforms. The theme of the work was clashing perspectives on creation of life, as seen by three mad scientists (performed by the three authors). The work featured a variety of new interfaces: a computer vision surface, an Arduino-based~\cite{ArduinoWebsite} heartbeat sensor and a network for communicating data from the interfaces to each of the performance computers.

These interfaces were built specifically for \textit{Vital LMTD} to
solve an artistic problem. We wanted to feature each performer as a
character on stage rather than hiding them behind their instruments
and computers. The only way to achieve this was to build interfaces
and a reliable performance environment so that the musical and visual
elements could be controlled away from a computer and as a by product
of the in-character actions of each performer.

\begin{figure}
\centering
\includegraphics[width=\columnwidth]{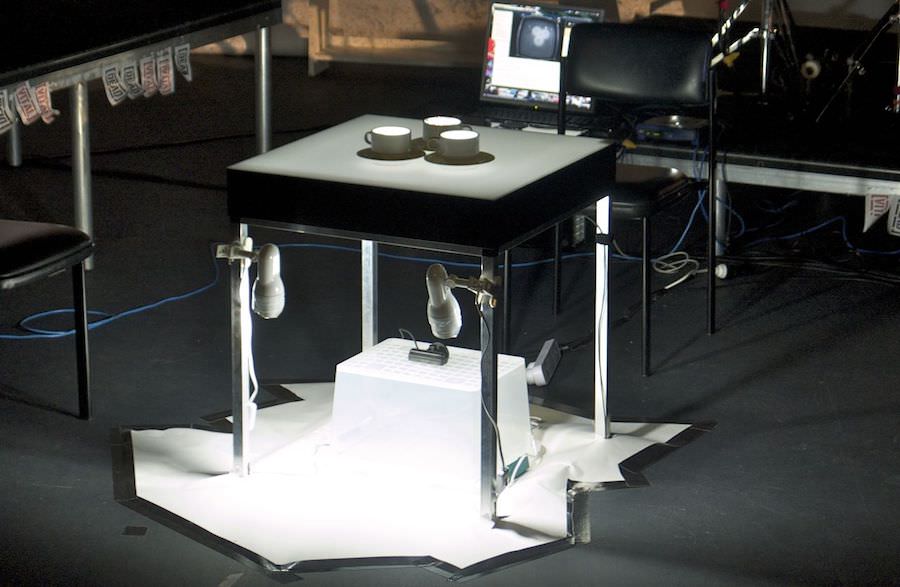}
\caption{The computer vision surface with fiducial-tagged teacups in \textit{Vital LMTD}}
\label{ComputerVisionSurface}
\end{figure}

\begin{figure*}
\centering
\includegraphics[width=\textwidth]{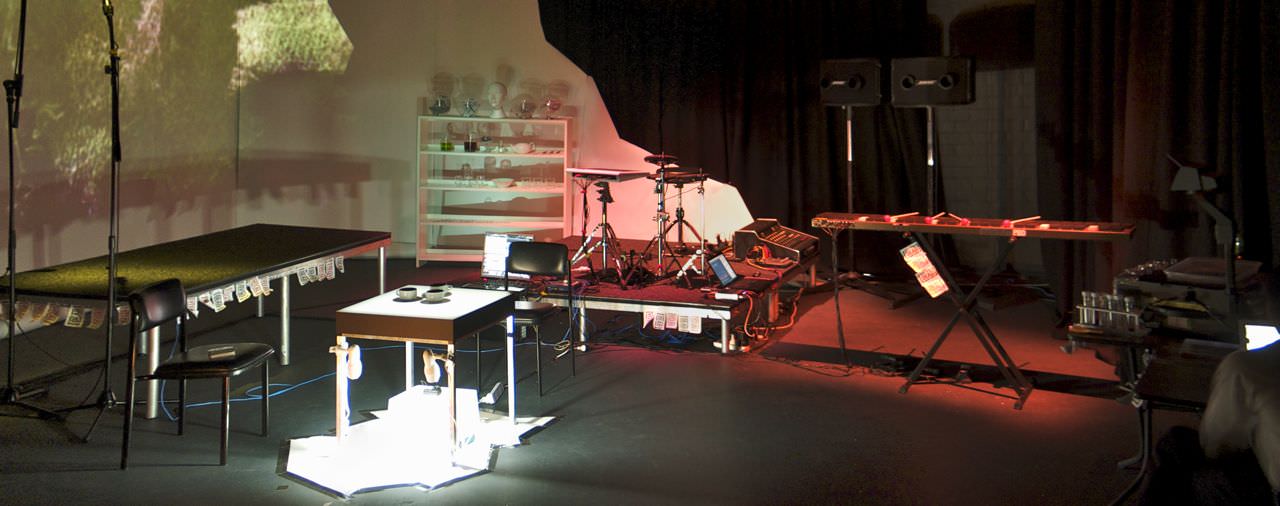}
\caption{The set of \textit{Vital LMTD} at \textit{The Street Theatre}}
\label{VitalLMTDSet}
\end{figure*}

The interfaces were also designed to be appealing and meaningful to the audience. Our set design (figure \ref{VitalLMTDSet}) emphasised our equipment and drew the audience's attention to the computer vision surface at the centre. It was important to us that the audience could see the tools that were underlying our show.

Another focus of our development was simplicity and reliability. It was not acceptable for us to stop the show to reset computers or check settings. By keeping our interfaces as simple as possible we were able to perform with very few mishaps.

The end result was that much of our show was controlled by gestures that were intentional within our narrative, for example, setting up scientific glassware. These gestures were also interpreted by the interfaces which could trigger a musical and visual response.
 
\section{Computer Vision Surface}

\noindent To meet our goal of featuring each artist as a character on stage, we used a coffee table sized computer vision surface as the centrepiece of our show. A camera below the surface tracked fiducial markers\footnote{Special patterns that can be tracked by computer vision software.} which were attached to teacups, lab equipment and playing cards that were manipulated on the table by the seated performers.

The software for the table was reacTIVision~\cite{KaltenbrunnerBencinca-reactivision} with the video image captured by a modified Playstation Eye USB camera~\cite{playstationEyeCamera}. The camera was modified with a wide angle lens purchased from Peau Productions~\cite{PeauProductions-website}. reacTIVision sends data about the location of objects on the surface over a network using OSC~\cite{osc-nime2009} and TUIO~\cite{TUIO_KBBC05} protocols.

The surface itself was a frosted perspex lightbox and a frame constructed using an aluminium shelving system. This frame was light and easy to dismantle. Illumination was provided by two desk lamps clipped to the upper part of the frame and aimed down at paper which reflected and diffused light back up to the surface. This strategy eliminated hotspots on the glossy underside of the perspex surface. The frame did not have sides allowing the audience to see the camera and lights.

The Playstation Eye camera provided better than expected performance for its cost. The wide angle lens allowed good focus and a much wider view than we required. A crucial element of this setup was the software driver for the camera. We used the community written driver for Linux~\cite{KaswyPS3Driver}, which allowed control over exposure and contrast under Ubuntu 9.04. Using this driver, reacTIVision was easily able to identify fiducial markers right to the edge of the surface. The available drivers for Mac OS X~\cite{macam-website} were not as successful. Exposure and contrast controls were not available and so reacTIVision could not detect fiducials as reliably. A  development version (2009-09-25) of this driver does provide these features.

The computer vision surface was an artistic success as well as a technical one. Rather than using it as a ``controller" for synthesisers or musical loops, the surface allowed the performer's in-character manipulation of props to influence the music and visuals. This influence could be brought into focus, for example, when placing a playing card triggered a corresponding musical cue. It could also be more subtle, such as the positions of teacups on the surface affecting the relative positions of musical and visual elements. These connections bound the musical and visual elements of the show to the physical movements of the performers in an implicit way, in keeping with our goals.

\section{Physical Computing and Arduino}
\begin{figure}
\centering
\includegraphics[width=\columnwidth]{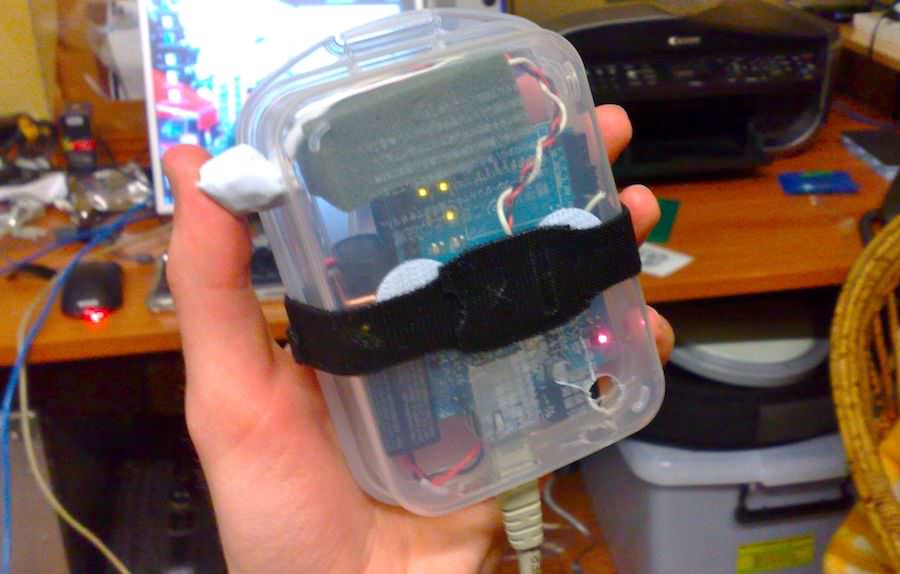}
\caption{Our networked heartbeat sensor in action.}
\label{HeartbeatSensorAction}
\end{figure}

\noindent One section of \textit{Vital LMTD} used a heartbeat sensor built using the Arduino electronics platform~\cite{ArduinoWebsite}. The interaction was simple - one performer applied the ``life sensor" to another and on each heartbeat an OSC message was sent over the network which triggered a sonic and visual response. 

The Arduino is a development board for creating smart electronic devices. It consists of a programmable microchip with supporting electronics as well as a  programming language and development environment. The Arduino board can be extended with ``shields", other boards that stack on top of it. We used the ethernet shield to provide network connectivity. With this, our sensor could send an OSC message to the performance computers each time it detected a heartbeat.

\begin{figure}
\centering
\includegraphics[width=\columnwidth]{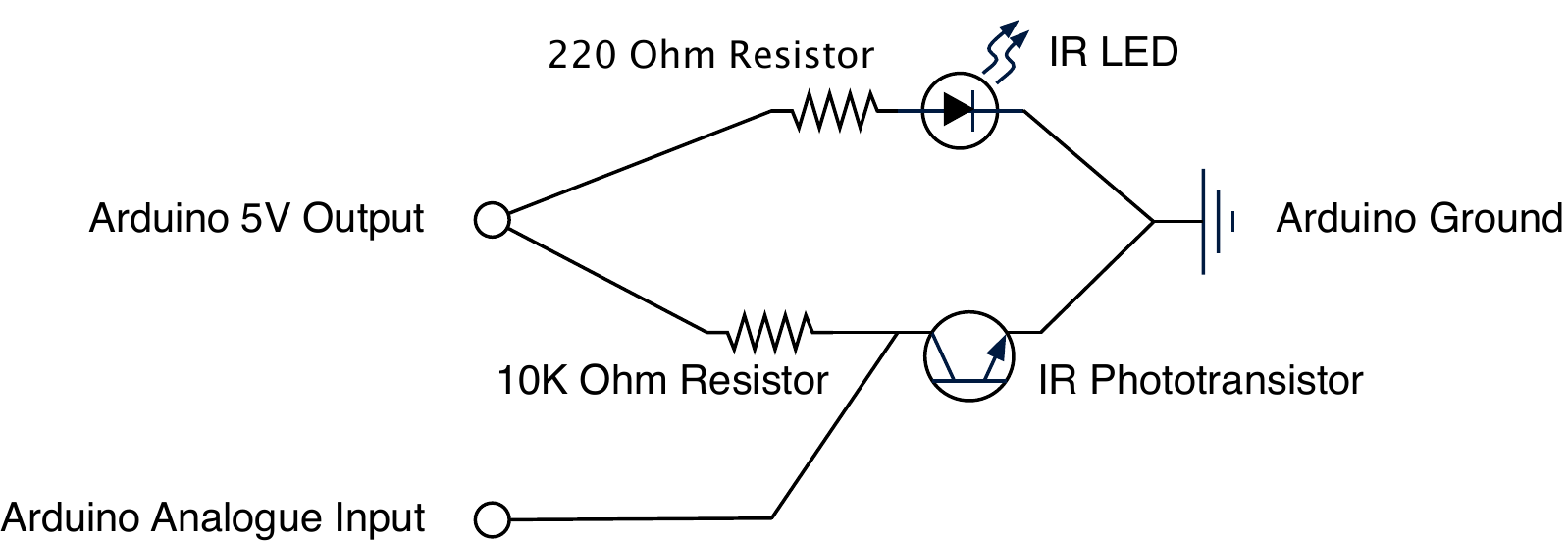}
\caption{Circuit for the heartbeat sensor.}
\label{HeartbeatSensorCircuit}
\end{figure}

The heartbeat sensor itself (figure \ref{HeartbeatSensorCircuit}) is simple to implement for the Arduino. The sensor consists of an IR LED, an IR detecting phototransistor and two resistors~\cite{GordonMcComb-RobotBonanza}. The IR LED and phototransistor are positioned next to each other and their lenses pressed against the skin of a finger. The sensor functions on the principle that the changing blood pressure in small blood vessels changes the opacity of skin. The Arduino tracks the brightness of light transmitted through skin and can detect heartbeats by identifying sharp increases in opacity (corresponding to more oxygenated blood). The Arduino code is available on Charles Martin's website~\cite{Martin-ArduinoHeartbeatSensor}.

\begin{figure}
\centering
\includegraphics[width=\columnwidth]{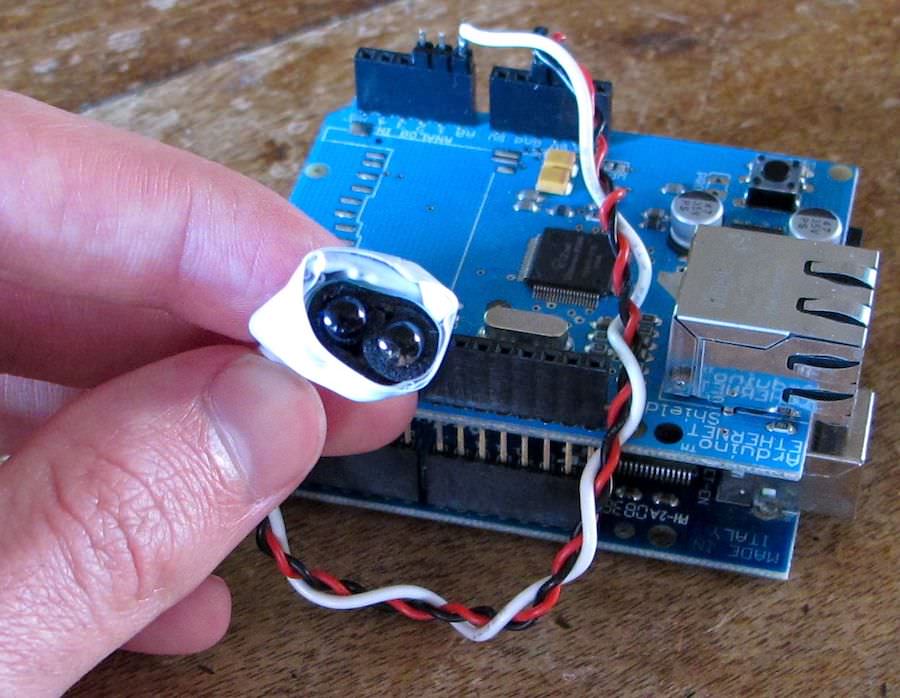}
\caption{The two lenses of the IR LED and Phototransistor.}
\label{HeartbeatSensorDetail}
\end{figure}

The Arduino was an easy system for building this kind of interface. The Ethernet Shield is well-documented and a library~\cite{recotana-OSCLibrary} is available for sending and receiving OSC messages. Our experience demonstrates that a battery powered Arduino sending OSC to all the performance computers is reliable, it is also more elegant than tethering a Arduino or other programmable system to a computer and sending OSC from there.

There are other examples of IR heartbeat sensors on the web including another  using Arduino~\cite{MengLi-StillAlive} and one that sends data over MIDI~\cite{Recotana-HeartbeatMidiController}. Both of these sensors featured an amplifier chip to boost the signal from the phototransistor. Our experience shows that this may not be needed. Our method of detecting a heartbeat, looking at the sum of recent changes in opacity, was accurate enough to use in performance without an amplifier and did not require calibration for different users.

In our show, the heartbeat sensor became a ``life sensor" a tool that can distinguish between the living and the dead. This prop was an interesting narrative device as well as a useful interface, further integrating our physical, musical and visual performance.

\section{Networked Performance}
\noindent Using a wired router on-stage allowed the computer based soundtrack and projected visuals to be synchronised and controlled by the various interfaces. The common language of Open Sound Control was used to facilitate these links. 

Our computer vision surface, heartbeat sensor and computers were configured to send OSC to our router's broadcast address. This meant that messages were sent to each client on our network, i.e. every performance computer. Towards the end of our development, we added extra devices such as a Nintendo Wii Remote sending OSC through the OSCulator\footnote{\url{http://www.osculator.net/}} software  and messages from MIDI percussion instruments converted to OSC using Pure Data. 

Since these devices sent data to all of our computers we could add and subtract functionality without having to alter the configuration of each device. It would be correspondingly easy to add another client to our performance network, listening to data from each interface.

\section{Conclusions and Future Work}
\noindent The networked interfaces featured by \textit{Last Man to Die} in \textit{Vital LMTD} allowed novel connections between our artforms. Most interesting to the authors was how the interfaces could be used as narrative features of our show and, in doing so, made more interesting to the audience.

It was an artistic and performative challenge to design interactions that avoided arbitrary movements for controlling our interfaces. In our early rehearsals we would move computer vision tracked teacups just to experience the audio and visual response. Later, we limited our interactions to those that made sense from our characters. This limitation made our interfaces meaningful as tools for gathering data from a sensible physical performance rather than as ``controllers" for music and visuals.

Currently, our interfaces do not provide feedback directly to the user, other than their influence on the projected visuals or music. The success of the networked heartbeat sensor has encouraged us to explore a networked wearable interface, that can provide feedback via vibrating motors to influence a performance implicitly. Projection mapping on our props and bodies in performance could also allow localised feedback for each performer.

\section{Acknowledgments}
The authors are supported by the A.C.T. Government, \textit{The Street Theatre}, Canberra, \textit{Belconnen Community Services}, Canberra, \textit{Canberra Youth Theatre}, Dr. Alistair Riddell and Gary France of the Australian National University.

\bibliographystyle{abbrvurlfornime}
\bibliography{cpm-2010-NIMEpaperBib}
\balancecolumns 
\end{document}